\documentclass{mn2e}
\usepackage{graphicx}

\newcommand{\beq}{\begin{equation}}
\newcommand{\eeq}{\end{equation}}

\begin{document}

\title{The Star Formation Law in a Multifractal ISM}

%
\author[K. Tassis]
{Konstantinos Tassis$^{1,2}$\\
$^{1}$Department of Astronomy and Astrophysics, 
The University of Chicago, Chicago, IL 60637, USA\\
$^{2}$The Kavli Institute for Cosmological Physics, 
The University of Chicago, Chicago, IL 60637, USA \\
}
\maketitle

\label{firstpage}
\begin{abstract}

The surface density of the 
star formation rate in different galaxies, as well as in different
parts of a single galaxy, scales nonlinearly with the surface density
of the total gas. This observationally established relation is known as the
Kennicutt-Schmidt star formation law. The slope of the star formation law has
been shown to change with the density of the gas against which the
star formation rate is plotted. This dependence implies a nonlinear
scaling between the {\em dense gas} and the {\em total gas} surface densities
within galaxies. Here, we explore a possible  interpretation of this scaling 
as a property  of the geometry of the interstellar medium (ISM), and
we find that it arises naturally if the topology of the ISM is
{\em multifractal}. Under the additional assumption that, at
very high densities, the star formation timescale is roughly constant,
the star formation law itself can also be recovered as a consequence
of the multifractal geometry of the ISM. 
The slope of the scaling depends on the width of
the global probability density function (PDF), and is between
$1.5$ and $1.6$ for wide PDFs relevant to high-mass systems, while it is
higher for narrower PDFs appropriate for lower-mass dwarf galaxies, in
agreement with observations. 

\end{abstract}


\section{Introduction}

The dependence of the star formation rate on gas density on large
scales is a subject of intense observational and theoretical
investigation, as it is both an important input for models of galaxy
formation and evolution and chemical evolution calculations, as well
as a critical test for theories of interstellar medium evolution and
star formation.  
Let us define the surface density of the star formation rate,
$\dot{\Sigma}_{\rm SF}$, as the mass of gas being converted into stars per
unit time per unit surface area, and 
$\Sigma_{\rm gas}$ as the total gas mass per unit surface
area. 
The two quantities have been found to obey a non-linear, power-law relation:
\begin{equation}\label{law}
\dot{\Sigma}_{\rm SF} \propto \Sigma_{\rm gas}^{n_{\rm gas}}\,, 
\end{equation}
where  
$n_{\rm gas} \approx 1.5$ for a large range of surface densities and system
morphologies. This correlation is known as the Kennicutt-Schmidt law
of star formation (Schmidt 1959, Kennicutt 1989).
The star formation law has been established through observations of the
global star formation and gas density in different galaxies (e.g.  
Kennicutt 1998, Misiriotis et al.\ 2004, Komugi et al.\ 2005); of
local gas and star formation densities in different galaxies (e.g.  
Wong and Blitz 2002, Boissier et al\ 2003);
and of local gas and star formation densities within a single galaxy (e.g. 
Misiriotis et al.\ 2006 in the case of the Milky Way, Schuster et al.\ 2007 in the case of M51).

Traditionally, this correlation has been interpreted in the literature
as a result of the density dependence of the star-formation
timescale. 
The star formation rate density can be expressed as a ratio of
the density of the gas available for star formation over a timescale
relevant to the conversion of the available gas to stars. 
Examples of such timescales that have been suggested in this 
framework are the free-fall timescale,  the turbulence crossing 
time over the scale height of the galaxy, the collapse timescale 
of large expanding shells with low Mach number
 (see e.g. Elmegreen 2002b and references therein), the orbital
 timescale of the galactic disk (Silk 1997), and
 the timescale for gas accumulation along magnetic flux tubes 
parallel to spiral arms, into the valleys created by the magnetic 
Rayleigh-Taylor (Parker) instability (see e.g. Shu et al. 2007). These
timescales {\em all} scale as the inverse square root of density, resulting in
an overall scaling of the star formation density as gas density to the
$1.5$. A more elaborate treatment based on the same principle was
presented by Krumholz and McKee (2005). This 
interpretation of the star formation law is very tempting 
because it is conceptually simple and elegant, and because it gives
the same result for several different processes which may be
controlling the star formation timescale. 
 However, a series of more recent observations have indicated
that this simple picture may not be a complete interpretation of the
  star formation law. 

First of all, the value $n_{\rm gas}\approx 1.5$, although consistent with
observations of a large range of star-forming systems, is by no means
unique. Appreciable scatter exists that cannot be entirely attributed
to observational uncertainties and cannot be comfortably reconciled
with  a theoretically set single value of $n_{\rm gas}$.  
For example, 
if only gas with local volume densities representative of molecular 
cloud cores is included in the gas surface density
(as is the case when  
tracers sensitive to higher 
density gas are used, such as HCN)
the resulting scaling is closer to linear, 
\begin{equation}\label{SFRvsDense}
\dot{\Sigma}_{\rm SF} \propto \Sigma_{\rm den, gas}^{n_{\rm dense}}\,, 
\end{equation}
where $\Sigma_{\rm den, gas}$ is the
surface density of the denser gas, and $n_{\rm dense}$ is now closer to $1$
rather than $1.5$ (e.g. Gao and Solomon 2004, Wu et al.\ 2005).  
Observations of dwarf
 galaxies on the other hand 
seem to favor much steeper scalings (values of
 $n_{\rm gas}$ much larger than $1.5$).  
For example, de Blok and Walter (2006) find for NGC 6822 a slope 2.04; Heyer et al.\ (2004) find for M33 a much steeper slope,
equal to 3.3;  in the case of IC10, Leroy et al.\ (2006), find a slope 
of $1.4$ for molecular gas and a slope much steeper
than $1.5$ for the total gas density (see their Fig. 14).

In addition,
since star formation is an inherently local phenomenon, the connection
between the small scales ($\sim 0.1{\rm \, pc}$) where
stars form, and the large scales ($\sim 1 {\rm \, kpc}$)
that the gas density observations and the associated timescales refer
to, may be quite complicated. 
The traditional
interpretation implicitly assumes that a meaningful mean density can
be assigned to the large scales where $\Sigma_{\rm gas}$ is
determined. However, this need not be the case. 
As a counter-example, we can consider the
situation where several dense objects are
sparsely distributed over a region of $\sim {\rm \, kpc}$
dimensions. The mean density we would derive would be
rather low, but it would not be representative
of {\em any} object in the region. Consequently, the
star-formation timescale calculated at that mean density would not
represent the real timescale over which any of the objects in the
region evolve.
Instead, the relevant timescale for star
formation should be constructed out of the densities of the individual
star-forming sites rather than the large-scale mean density. 

Finally, the interstellar medium (ISM) is a highly complex
system, and its structure and properties are the result of a synthesis
of nonlinear phenomena and instabilities, not necessarily connected
with one another. Star formation on the other hand only occurs at the
density peaks of the ISM, and it is likely that the star
formation law may be conveying information about 
the connection between the large-scale low-density ISM and its
small-scale high-density peaks, rather than information about the star
formation process itself. 

The connection between the
high-density and low-density ISM gas is imprinted in different
incarnations of the star formation law. Observations have established that the
slope of the scaling of the star formation rate surface density with
the gas surface density depends on the minimum density of the gas accounted
for in observations [Eqs. (\ref{law}) and (\ref{SFRvsDense})]. 
If now we demand that Eqs. (\ref{law}) and (\ref{SFRvsDense}) {\em
  both} hold, this
implies a scaling between the surface densities of total gas and dense
gas, 
\begin{equation}\label{densdens}
\Sigma_{\rm den, gas} \propto \Sigma_{\rm gas}^n\,,
\end{equation}
with $n=n_{\rm gas}/n_{\rm dense} \approx n_{\rm gas} \approx 1.5$.
 
In this paper we investigate this connection, and we examine the
extent to which such a nonlinear scaling of the dense gas
surface density with the total gas surface density can be interpreted
 as a topological property of the ISM. 
In particular, we investigate whether such a scaling can arise
naturally in
{\em multifractal} topologies. 

While fractal structures are characterized by
a {\em unique} non-integer generalized dimension, a {\em spectrum} of
generalized fractal dimensions is required to fully describe
multifractal structures. In a multifractal structure, we can define a
scaling exponent which characterizes the structure locally, around
each point. Sets of points that share the {\em same} scaling exponent
constitute a fractal. The distribution of fractal dimensions of all
such fractals is the multifractal spectrum, which describes the
distribution of geometries present in a complex structure.

The gas and dust in the ISM is organized in complicated structures
with irregularities present over a wide range of scales. Different
studies have used different techniques to quantify the properties of
these structures and describe the geometry of the ISM:
e.g., autocorrelation function (Dickman and Kleiner 1985), power
spectra (St\"{u}tzki et al. 1998); perimeter/area relation (Bazell and
Desert 1988; Scalo 1990; Vogelaar and Wakker 1994; Westpfahl et al.\
1999 in HI for members of the M81 group), wavelet analysis (Langer et
al. 1993), principal component analysis (Heyer and Schloerb 1997),
multifractal scaling (Chappell and Scalo 2001). These studies  
find that structure in the ISM has a scale-free, hierarchical
appearance. This complexity over a wide range of scales has lead to the
proposition that a multifractal ISM geometry with a lognormal
probability density function (PDF)
is consistent with many of the observed features, including the mass
and size distribution of clouds and  the ISM filling factor 
(Elmegreen 2002a;  
Elmegreen et al.\ 2006). The physics is usually associated with
compressible turbulence (e.g., Mac Low and Klessen 2004, Scalo and
Elmegreen 2004), however the multifractal scaling is much more
generic and may arise from many different multiplicative hierarchical
processes (e.g. 
Colombi, Bouchet, and Schaeffer 1992; 
Sylos Labini and Pietronero 1996).

It should be noted that 
the ISM cannot be strictly scale-free, since
optical observations of a variety of galaxies suggest  an upper
characteristic scale ($\sim$1 kpc) that can either be the Jeans
length (e.g. Elmegreen 2004) or the wavelength of the Parker
instability (Mouschovias, Shu and Woodward 1974). A lower characteristic scale
also exists, at the order of $\sim$0.1 pc (Blitz and Williams 1997;
Barranco and Goodman 1998; Goodman et al. 1998). However, in between
these scales, a variety of different processes and instabilities can
lead to multifractal scalings and lognormal PDFs (e.g. Vazquez-Semadeni 1994; Kravtsov 2003; Wada \& Norman 2007).

 The purpose of this work is to examine whether a nonlinear scaling of
 the dense gas 
surface density with the total gas surface density similar to that of
Eq. (\ref{densdens}) can arise naturally in multifractal
ISM geometries, whether it is robust enough to resemble the
persistence and uniformity of $n_{\rm gas}/n_{\rm dense}$ 
over a variety of star forming systems with different local
conditions, and whether it 
can allow for different slopes seen e.g. in observations of
dwarf galaxies under physical conditions similar to those found in
these systems.  

A geometrical explanation to
Eq. (\ref{densdens}) has an additional tantalizing 
potential consequence.  
Equation (\ref{SFRvsDense}) implies that the star formation rate
surface density scales linearly with the very dense gas, or,
equivalently, that at those high densities the star formation
timescale is roughly constant and independent of $\Sigma_{\rm den,gas}$: 
\begin{equation}\label{smallscale}
\dot{\Sigma}_{\rm SF} = \frac{\Sigma_{\rm den, gas}}{\tau_{\rm SF, dense}}
\end{equation}
with $\tau_{\rm SF, dense}\sim {\rm const}$. 
However, in this case $\dot{\Sigma}_{\rm SF} \propto \Sigma_{\rm den,
  gas}$ and if $\Sigma_{\rm gas} \propto \Sigma_{\rm den,gas}^n$ due
to geometry, then
$\dot{\Sigma}_{\rm SF}\propto \Sigma_{\rm gas}^n$. The observationally
motivated assumption of a star formation timescale independent of
$\Sigma_{\rm den}$ at high
densities, in combination with a topologically-driven scaling between
dense and total gas, can thus provide a geometrical interpretation of the star
formation law itself.

This paper is structured as follows. Our formulation and our
algorithms for the reproduction of multifractal three-dimensional
geometries are discussed in Section \ref{form}. Our results are
presented in Section \ref{theres}. We discuss our findings in 
Section \ref{disc}, and we summarize our conclusions in Section \ref{sum}. 

\section{Formulation}\label{form}

In order to 
investigate how the mean surface density of very dense gas
correlates with the surface density of the total gas in different multifractal ISM topologies, we need to
generate a multifractal structure, and perform appropriate
``observations'' of its total and dense ``gas'' surface densities. In
this section, we describe these procedures. 

To generate a multi-fractal, we adapt the algorithm of
Borgani et al. (1993), which is a modification of the $\beta$ model
(Frisch et al. 1978) and the random-$\beta$ model (Benzi et
al. 1984) of fully developed turbulence. We start with a parent
three-dimensional cube of side $L$, which we divide into $2^3$ equal-volume
subcubes, each of which inherits some fraction $f_i$ of the parent-cube
mass, where $i=1,..,8$. Mass conservation implies that $\sum _{i=1}^{8} f_i
= 1$. We repeat the fragmentation (where each subcube now becomes a
parent-cube)  $H$ times. The fraction of the total mass contained in
each final cube of volume $L^3/2^{3H}$ depends on its fragmentation
history. 

The properties of the final structure depend on the choice of
$f_i$. If all $f_i$ are non-zero and equal, then the result is a
homogeneous structure. If some of the $f_i$ are zero and all of the
non-vanishing $f_i$ are equal, then the result is a monofractal
structure, with a fractal dimension uniquely set by the choice of
$f_i$. If the non-vanishing $f_i$ are not equal, then a multi-fractal
structure results. To obtain a pattern-free structure that still
retains  the scaling properties of a fractal (self-similarity) 
or a multifractal (self-affinity), the pattern of $f_i$ is not kept
constant in all iterations: each $f_i$ value is assigned to a random
subcube $i$ in each iteration. 

Each multifractal cube is taken to represent one ISM topology. 
We investigate how the total gas surface density correlates with the
dense gas surface density in different parts of this object in the following way:
once a multifractal cube is constructed according to the algorithm
above, a line-of-sight is selected, parallel to one side of the
cube. The face of the cube which is perpendicular to that line is then
split into $2^{2(H-3)}$ square patches (each side of the face is split
into $2^{H-3}$ segments). For each of these patches, an
``observation'' is made of its total surface density, by summing up
all of the mass within the volume defined by the patch and extending
along the line of sight, and dividing by the surface area of the
patch. Similarly, the ``dense gas'' surface density is calculated by
summing up all of the mass within regions of local volume density
above the ``dense gas'' threshold within the same volume, and dividing by the
surface area of the patch.  

To test the robustness of our results against variations of the
detailed properties of the multifractal cube,
we have
repeated our measurements for seven distinct multifractals. The
properties of these multifractals ($f_i$ and number of refinement
levels, $H$) are given in Table \ref{thetable}.

\begin{table}
\caption{Mass fractions $f_i$ and levels of refinement $H$ for the seven multifractals investigated in our study. Multifractals are listed here in order of decreasing PDF width (see Fig.\ \ref{fig4}).\label{thetable}}
\begin{tabular}{lccccccccc}
\hline
& $H$ & $f_1$ & $f_2$ & $f_3$ & $f_4$ & $f_5$ & $f_6$ & $f_7$ & $f_8$ \\
\hline 
A & $7$ & $0.6$ & $0.2$ & $0.1$ & $0.05$ & $0.04$ & $0.005$ & $0.004$ &  $0.001$\\ 
B & $6$ & $0.6$ & $0.1$ & $0.1$ & $0.1$ & $0.09$ & $0.005$ & $0.004$ & $0.001$\\ 
C & $5$ & $0.6$ & $0.2$ & $0.1$ & $0.05$ & $0.04$ & $0.005$ & $0.004$
& $0.001$\\ 
D & $6$ & $0.44$ & $0.17$ & $0.1$ & $0.1$ & $0.1$ & $0.05$ & $0.03$ & $0.01$\\
 E & $6$ & $0.4$ & $0.15$ & $0.1$ & $0.1$ & $0.1$ & $0.05$ & $0.05$ & $0.05$\\ 
F & $7$ & $0.3$ & $0.1$ & $0.1$ & $0.1$ & $0.1$ & $0.1$ & $0.1$ & $0.1$\\ 
G & $6$ & $0.3$ & $0.15$ & $0.1$ & $0.1$ & $0.1$ & $0.1$ & $0.1$ & $0.05$\\ 
\hline
\end{tabular}
\end{table}

\section{Results}\label{theres}

The first questions we seek to answer are whether the scaling between
the very dense gas 
and total gas in a multifractal medium resembles that of
Eq. (\ref{densdens}); and  whether such a scaling is robust enough
against variations in the detailed properties of the multifractal
(parameterized in our model by the values of the $f_i$ and $H$) so that
such a scaling can be viewed as an intrinsic property of multifractal
geometries, rather than as a result of fine-tuning of the 
multifractal model. 

\begin{figure}
\includegraphics[width=3.0in]{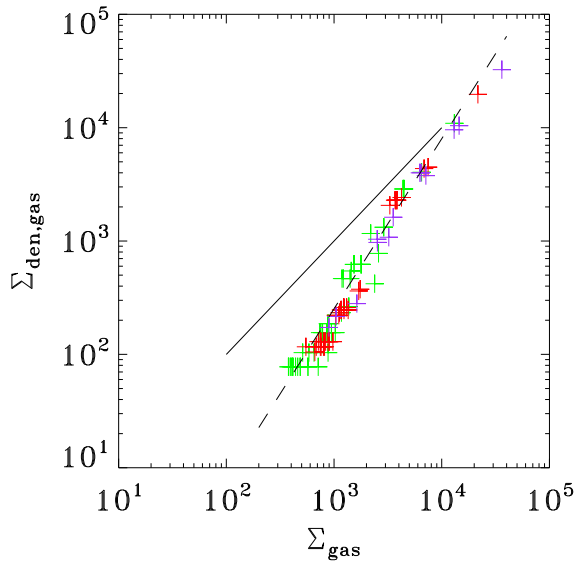}
\caption{Scaling of dense gas surface density vs. total gas surface density  for
  multifractals A (green), B (red), and C (purple).  The solid and dashed
  lines correspond to the $\Sigma_{\rm den,gas} \propto \Sigma_{\rm
    gas}$ and $\Sigma_{\rm den,gas} \propto \Sigma_{\rm
    gas}^{1.5}$ scalings respectively.
}\label{fig1}
\end{figure}

We address these questions in Fig. \ref{fig1}, which
shows the surface density of dense gas, $\Sigma_{\rm
  den,gas}$, in this case defined as gas with local volume density
at least $3\times 10^3$ times the mean volume density, 
as a function of the surface density of total gas $\Sigma_{\rm
  gas}$, for three different multifractals (A,
B,
and C).
Each datapoint corresponds to a different patch
within the  multifractal cube. 
The three fractals depicted in this figure have
different mass fractions, $f_i$, however they all have ``wide'' global
PDFs (see upper panel of Fig. \ref{fig4} and
discussion below). 
In all cases, the scaling between $\Sigma_{\rm gas}$ and
$\Sigma_{\rm den,gas}$ is clearly nonlinear (significantly
deviates from the linear relation plotted in Fig. \ref{fig1} 
with the solid line), and has a scaling slope of $\approx 1.5$.

To understand the origin of the nonlinearity of the scaling, we need
to examine how the gas is distributed over different local densities.  
This distribution is quantified by the PDF, which is defined as the
distribution of volume fraction with respect to local density (where
the local density is  measured in units of the mean density of the
cube).  
In order to have a nonlinear scaling between the surface densities of total gas and gas density peaks, the PDFs of different patches within the cube {\em must} be different from each other. This requirement
is straight forward to prove.
Let the local PDF within a patch $i$ of volume $V_i$ be $g_i(\rho) = (1/V_i)dV/d\rho$.
Then, the total dense gas mass in patch $i$
is an integral of the PDF above some high density threshold $\rho_{\rm
den}$,
\begin{equation}
M_{\rm den,gas,i} = V_i\int_{\rho_{\rm den}}^\infty g_i(\rho)\rho d\rho\,,
\end{equation}
and the total gas mass of the same patch is an integral of the PDF over all densities, 
\begin{equation}
M_{\rm gas, i} = V_i\int_{0}^\infty g_i(\rho) \rho d\rho\,,
\end{equation}
while the ratio of surface densities
is given by 
\begin{equation}\label{seven}
\left(\frac{\Sigma_{\rm den, gas}}{\Sigma_{\rm gas}} \right)_i= 
\frac{M_{\rm den, gas,i}/A_i}{M_{\rm gas,i}/A_i} = \frac{\int_{\rho_{\rm
      den}}^\infty g_i(\rho) \rho d\rho}{\int_{0}^\infty g_i(\rho) \rho d\rho}\,,
\end{equation}
where $A_i$ is the surface area of the patch perpendicular to the line
of sight. If now  
the PDF remains the same among different patches (all $g_i$ are
identical), the right-hand side of Eq. (\ref{seven}) is always
constant, resulting in an identically linear scaling, $\Sigma_{\rm
  den,gas} \propto \Sigma_{\rm gas}$. A similar result is obtained, as
long as the PDF spread is sufficiently large, if
the $g_i$ have the same spread but different mean densities. 
A nonlinear 
scaling can {\em only} arise if different patches exhibit different
PDF spreads. This feature is an inherent property of multifractals, as
we show below.  
\begin{figure}
\includegraphics[width=3.0in]{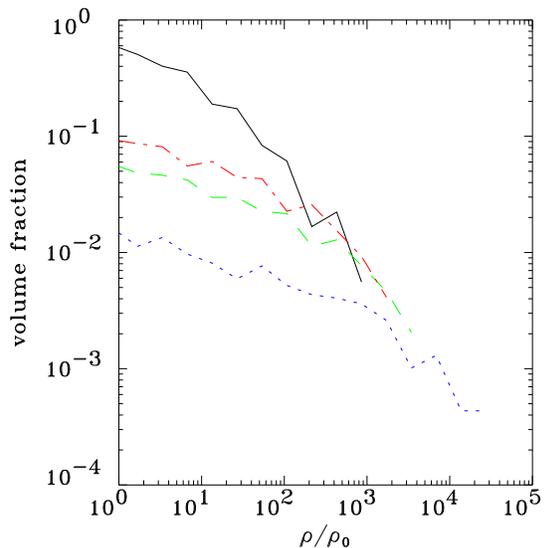}
\caption{Probability density functions of different patches  
of multifractal A. The total gas
surface density is lowest for the patch represented by the solid
black line, and progressively increases for the patches depicted in
dashed red, long-dashed green, and dotted blue respectively. 
}\label{fig2}
\end{figure}

Figure \ref{fig2} shows local PDFs for four different patches of varying
mean density for multifractal A. In our multifractal
geometry, the local PDFs are
not identical in spread to each other, and they are not identical to the global
PDF of the cube (plotted with the thin solid black line in the upper panel of
Fig. \ref{fig4}). Instead, they are truncated at different densities
depending on their total gas surface density. 
The total gas
surface density is lowest for the patch represented by the solid
black line, and progressively increases for the patches depicted in
dashed red, long-dashed green, and dotted blue respectively.  
The lower the 
total surface density of the patch, the lower the density at which the
PDF is truncated.  
It is this effect which 
is the origin of the
nonlinearity in the scaling of $\Sigma_{\rm den,gas}$ with
$\Sigma_{\rm gas}$. A similar
dependence of the local PDF on the mean density was pointed out by
Kravtsov (2003) in the case of cosmological simulations,  
where 
the star formation law was reproduced in large scales, despite
the fact that the local star formation recipe featured a timescale
{\em constant} with density.

We have demonstrated that a nonlinear scaling between very
dense and total gas surface densities 
arises naturally in 
multifractal geometries, and we
have traced its origin in the property of the local multifractal PDF
to extend to higher local densities in regions of higher mean density.
However, there are still several points to be addressed before we can 
draw conclusions on how robust this scaling really is, and how it
may relate to the star formation law. These points are the
following.
\begin{enumerate}
\item The definition of ``dense gas'' we used in our discussion of
  Fig. \ref{fig1} was rather arbitrary. We would like to understand 
how the slope of the scaling may depend on this definition. 
\item Although we have shown that different multifractals show similar
  behaviour, we would like to understand whether scalings with slope other than $\approx 1.5$ can also be recovered, and under what conditions. 
\item Although the detailed values of $f_i$, together with the number
  of refinement levels, uniquely determine the multifractal scaling
  properties, it would be desirable to identify a global
property of the multifractal, which controls the slope of the scaling.
\item Finally, we would like to examine 
whether the conditions under which scalings with slopes different that
$\approx 1.5$ appear in our simulated multifractals possibly 
  mirror situations in nature where the $n_{\rm gas}$ slope of the 
star formation law entering Eq. (\ref{densdens}) also has different
values (as, for example, in the case of dwarf galaxies).  
\end{enumerate}
We address these issues with the help of Figs. \ref{fig4} and \ref{fig3}.

\begin{figure}
\includegraphics[width=3.0in]{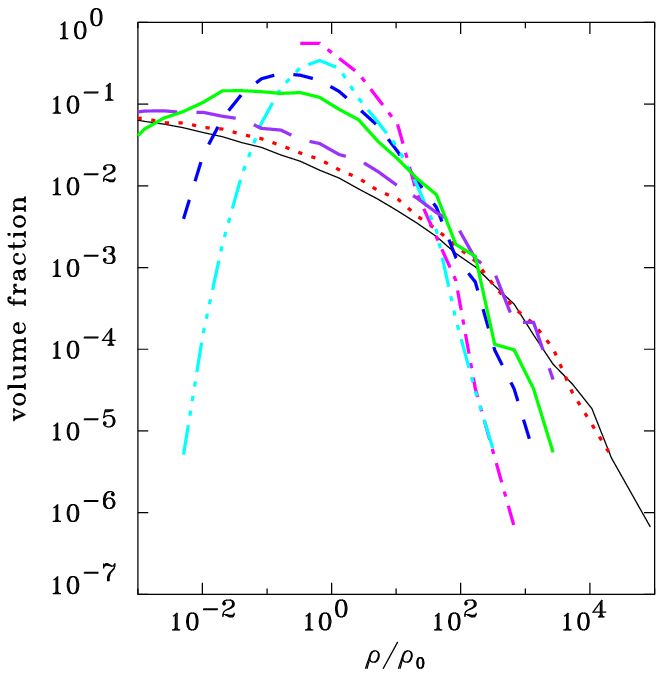}
\includegraphics[width=3.0in]{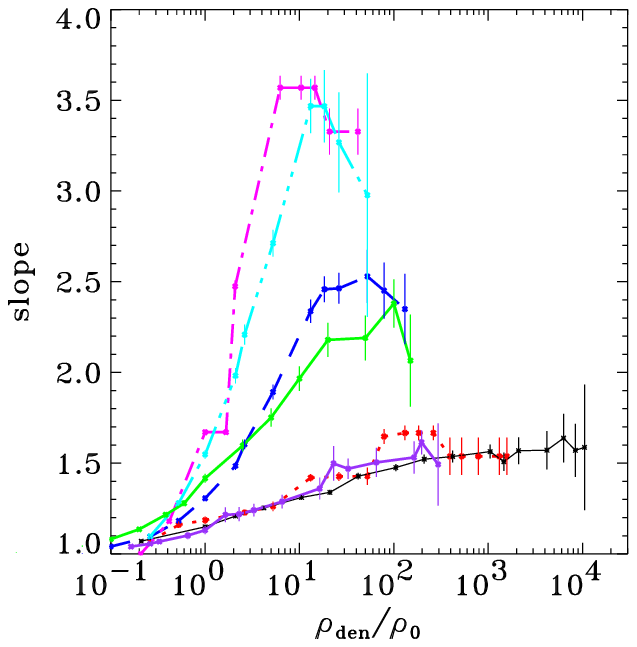}
\caption{Upper panel: probability density functions for each
  multifractal.
Different line styles and colours correspond to different
  multifractals as follows: thin solid black $\rightarrow$ A; dotted
  red $\rightarrow$ B; long-dashed purple $\rightarrow$ C; thick solid
  green $\rightarrow$ D; dashed blue $\rightarrow$ E;
  double-dot--dashed cyan  $\rightarrow$ F; dot-dashed magenta
  $\rightarrow$ G. Lower panel: dependence of the exponent $n$ of the
  scaling  
  between dense gas surface density and total gas surface density on
  the value of 
  ``dense gas'' density threshold (in units of the mean density of the
  multifractal); line styles and  colours as in
  the upper panel.
}\label{fig4}
\end{figure}
The upper panel of Fig. \ref{fig4} shows the PDF of the cube as a
whole for each multifractal described in Table \ref{thetable}. The
global PDF can be fitted well by a lognormal 
in all cases. The lower panel of the same figure shows the dependence of the
slope $n$ (the scaling between $\Sigma_{\rm den, gas}$ and $\Sigma_{\rm gas}$, see Eq. \ref{densdens}) on the value of the dense gas
density threshold, $\rho_{\rm den}$ in units of the mean density of
the cube, $\rho_0$. The points in the lower panel are generated through
fitting power laws to scatter plots similar to those of
Fig. \ref{fig1}, where $\Sigma_{\rm den, gas}$ is, in each case,
calculated with the appropriate definition of $\rho_{\rm den}$. The
error bars in the data points represent the $1\sigma$ uncertainty on
the slope of those fits. For higher values of the threshold $\rho_{\rm
  den}$,
fewer points are found within the cube which include mass at such
densities, and for this reason the uncertainty on the scaling slope
increases. 
From the two panels of Fig. \ref{fig4}, we can draw the following
conclusions.

\begin{figure}
\includegraphics[width=3.0in]{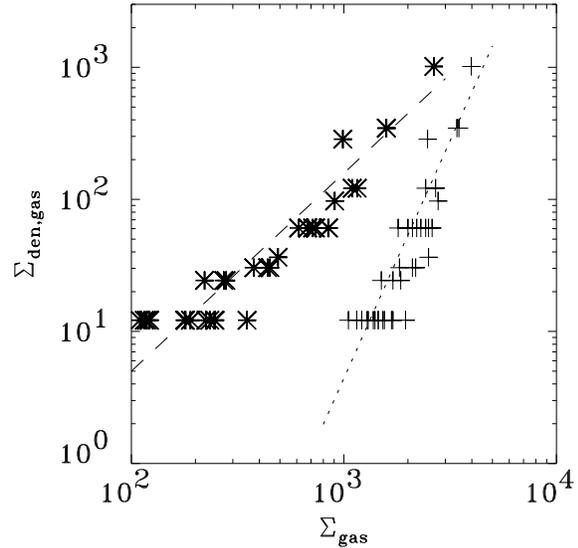}
\caption{ Dependence of scaling slope $n$ on the definition of
  $\Sigma_{\rm gas}$. Crosses: 
scaling of $\Sigma_{\rm den,gas}$ with $\Sigma_{\rm gas}$
for multifractal G. $\Sigma_{\rm den, gas}$ includes gas of local
volume density at least $26\rho_0$, and $\Sigma_{\rm gas}$  gas of any
local volume density. Stars: same scaling, but with $\Sigma_{\rm gas}$
now including only gas of  local volume density at least $5\rho_0$. 
The dashed and dotted lines are fits to the two scalings, 
and have slopes of $3.6$ and $1.5$ respectively.}\label{fig3}
\end{figure}

\begin{enumerate}
\item When $\rho_{\rm den}/\rho_0$ is very
  low, most of the gas mass in the cube is counted as ``dense gas'', and the slope $n$  of the  $\Sigma_{\rm den, gas}$-$\Sigma_{\rm gas}$ is close to unity. As $\rho_{\rm den}$ increases, the scaling slope also increases; however, this trend does not continue indefinitely. As long as ``dense gas'' is at least
  one order of magnitude denser than the mean ISM density,
the slope of the nonlinear scaling between  $\Sigma_{\rm den,
    gas}$ and  $\Sigma_{\rm gas}$  saturates to some value and no longer changes appreciably with increasing  $\rho_{\rm den}$. 

\item 
From the lower panel of
  Fig. \ref{fig4} we can also identify a trend of the saturation value
  of the slope to {\em decrease} as the highest density achieved in
  the multifractal cube (in units of the mean density) increases. The
  saturation value of the slope then settles at a value between
  $1.5-1.6$ for multifractals achieving increasingly high maximum
  local densities (e.g. multifractals A, B, and C, shown with 
the thin solid black line, red dotted line, and 
purple long-dashed line in Fig.\ \ref{fig4}). 

\item Although the maximum density achieved in the multifractal cube
  is a useful first tool to discern the trends in the slope
  of the $\Sigma_{\rm den, gas}$-$\Sigma_{\rm gas}$ relation, a better 
quantity that can be used to
  characterize the properties of the multifractal is
  the width of the global PDF. As can be seen
  by comparing the upper and the lower panels of Fig. \ref{fig4},  
  as long as the global PDF is ``wide'' (its full
width at half-maximum is roughly greater than 2 decs in
density), the slope saturates to a value between $1.5$ and  $1.6$, whereas
when the PDF is narrower, the scaling law slope steepens and $n$ saturates
to increasingly high values for increasingly narrow PDFs. 
Multifractal C (long-dashed purple) and multifractal D (thick solid
green) achieve comparable 
maximum local densities, however multifractal D has a considerably
narrower global PDF  and as a result $n$ saturates to higher values. 

\item Physically, a narrow PDF
corresponds to a ``puffy'' ISM, characteristic of lower-mass systems
with shallower gravitational potentials and shallower density
gradients
(e.g. Wada and Norman 2007; Tassis, Kravtsov and Gnedin 2006).
An increase of $n$ (or, equivalently, $ n_{\rm gas}$)  
in low mass systems such as the one seen
in Fig.\ \ref{fig4} is
consistent with observations of dwarf galaxies (e.g. de Blok
and Walter 2006; Heyer et al.\ 2004; Leroy et al.\ 2006).

We discussed above the dependence of $n$ on the
density threshold $\rho_{\rm den}$ relevant to $\Sigma_{\rm
  den,gas}$. We would now like to address how $n$ changes if
$\Sigma_{\rm gas}$ no longer represents the total gas surface density,
but only includes {\em part} of the gas, and more
specifically gas of density higher than some threshold
$\rho_{\rm gas}$.
There are observational indications that the value of $n_{\rm gas}$ (and
thus of $n$) depends on $\rho_{\rm gas}$.
 For example, observations of dwarf galaxies indicate
that when the star formation surface density is plotted against the
surface density of denser (e.g. molecular) gas, rather than against
the total gas surface density, the scaling has a smaller slope
 (e.g. Heyer et al.\ 2004, Leroy et al.\ 2005, 2006). There is a
 simple intuitive 
 explanation for this result. As $\rho_{\rm gas}$ increases, the
 densities of the gas included in $\Sigma_{\rm gas}$ and  
$\Sigma_{\rm den,gas}$ become increasingly similar, and so the scaling
 shifts to slopes closer to linear.  In the extreme
 case where $\rho_{\rm gas}$ becomes equal to $\rho_{\rm den}$, 
the scaling between the two
 quantities becomes identically linear. 

We explicitly demonstrate that this trend is reproduced in multifractal ISM
geometries in Fig. \ref{fig3}. We plot, for multifractal G (a
multifractal with a ``narrow'' PDF, appropriate for dwarf
galaxies), $\Sigma_{\rm den, gas}$ against $\Sigma_{\rm gas}$ in two
different ways. In the first case (depicted by crosses), gas of all
densities is included in $\Sigma_{\rm gas}$ ($\rho_{\rm gas}$ is
zero). The best-fit power law to 
this scaling has a slope of $3.6$. In the second case, only gas of
volume density higher than five times the mean density in the cube is
included in $\Sigma_{\rm gas}$ ($\rho_{\rm gas} = 5\rho_0$). As
expected, the best-fit power-law to 
this scaling is shallower, and has a slope of $1.5$. In both cases,
the density threshold for inclusion in $\Sigma_{\rm den,gas}$ is 26
times the mean density of the cube. We should note that the two threshold
densities for $\Sigma_{\rm gas}$ are not special and are simply used
as examples of the intuitively expected trend of the slope.
\end{enumerate}

\section{Discussion}\label{disc}

It is noteworthy that a geometrical interpretation for
the $\Sigma_{\rm den,gas} - \Sigma_{\rm gas}$ scaling has an
interesting consequence for the star formation law
itself. Observations of $\dot{\Sigma}_{\rm SF}$  suggest that the
scaling of $\dot{\Sigma}_{\rm SF}$ with $\Sigma_{\rm den,gas}$ is
linear (Gao \& Solomon 2004, Wu et al.\ 2005), which implies that the
star formation timescale for very high density gas is roughly
constant (see Eq. \ref{smallscale}) and independent of $\Sigma_{\rm
  den, gas}$. This however in turn implies that
$\dot{\Sigma}_{\rm SF}$ scales with $\Sigma_{\rm gas}$ in the same way
as $\Sigma_{\rm den,gas}$:
\begin{equation}
\dot{\Sigma}_{\rm SF} \propto \Sigma_{\rm gas}^n\,.
\end{equation}
Thus, if we assume that (a) the ISM has a multifractal topology and
(b) the star formation timescale at high densities is roughly
constant and does not depend on $\Sigma_{\rm den,gas}$, we
arrive to a geometrical interpretation of the star 
formation law. It is tantalizing  that if we do accept
these two assumptions, many properties of the star formation law are
recovered naturally. The robustness of the $1.5$ slope for many
different systems is the result of $1.5$ being the saturation value of
$n$ for all wide global PDFs, which are expected for high-mass
systems. The steepening of the star formation law in dwarf galaxies
(de Blok and Walter 2006; Heyer et al.\ 2004;   Leroy et al.\ 2006) is
the result of a higher saturation value of $n$ for narrower PDFs,
appropriate for low-mass systems  (Wada \& Norman
  2007). The dependence of the star formation
law slope on the tracer used to determine $\Sigma_{\rm gas}$, and the
decrease of the slope for tracers sensitive to denser gas   (e.g.\
Heyer et al\ 2004, Leroy et al.\ 2005; 2006), is 
the result of the dependence of $n$ on the density threshold of
gas included in $\Sigma_{\rm gas}$, and its decrease when this
threshold increases (see discussion of Fig.\ \ref{fig3}).

Thus, in this framework, both the universality of the
law, as well as deviations from it under certain conditions and in
specific systems, have a uniform interpretation as effects of the
geometry of the ISM. Because multifractal geometries are the
ubiquitous outcome of the combined effects of a variety of nonlinear
processes such as the ones shaping the scalings of the ISM, this
interpretation is not based on any assumption about {\em which} energy
inputs and dynamical processes dominate in ISM physics. Conversely,
reproduction of the star formation law by any ISM model  cannot be
on its own regarded as an indication that the model is a complete or
accurate description of ISM dynamics; rather, it indicates that the
model succeeds in reproducing a multifractal geometry with a wide
enough PDF from which the star formation law comes about as a natural
outcome.

The connection between a lognormal PDF and the star formation law has
been suggested by  
Elmegreen (2002b), however in this case
a complicated dependence of the efficiency on density above the
threshold was also introduced, in contrast to our work, where no such
dependence is either assumed or found to be necessary. 
Results from cosmological simulations (Kravtsov 2003) also point in the
direction of the star formation law arising from lognormal PDFs with a
constant small-scale star formation timescale, consistent with the
picture suggested by the observations of Gao \& Solomon (2004) and Wu
et al.\ (2005). 

From a theoretical point of view, a roughly constant timescale for
high-density gas can be attributed to star formation
being a threshold phenomenon, in the sense that stars form only out of
gas in the high-end tail of the PDF which exceeds some threshold in local volume
density. Observationally, this concept is supported by 
studies of the local environments of protostars, which are always
found in the densest parts of molecular clouds, the dense molecular
cloud cores (e.g. Enoch et al.\ 2007). 
Once the local volume density exceeds some value such
that gravity overtakes the forces supporting the star-forming cloud,
the dense gas collapses to form stars within a timescale corresponding
to the threshold density.
The nature of the relevant threshold and
associated timescale depends on the relative dynamical 
importance of different forces in star-forming clouds.
For example, in magneticly supported
clouds, the threshold (that needs to be reached through ambipolar
diffusion) is the critical mass-to-flux ratio, while the relevant
collapse timescale once the threshold is exceeded is the magnetically
diluted dynamical timescale (see e.g. reviews by
Mouschovias 1987, 1996 and references therein). If
magnetic fields are not dynamically important, the relevant timescale
(free-fall time) is very similar (see e.g. Vazquez-Semadeni et al.\ 2005 and
references therein). Although these timescales 
depend on local density, they have a fixed value $\tau_{\rm
  SF, dense}$ at the threshold, and this may be the origin of the Gao
and Solomon (2004) result. 

Even if the detailed local conditions in individual molecular clouds
(magnetic field, Mach number distribution) differ so that the threshold
density, as well as the associated star formation timescale at the
threshold, differ as well, the star formation law will not be
affected as long as the variations
are not too large . The threshold density for star formation is determined by
the structure of the molecular cloud cores (at scales of
$\sim 0.1 {\rm \, pc}$). On the other hand, $\Sigma_{\rm den,gas}$ is
determined by the masses, numbers, and {\em distributions} of these
cores within their large-scale environment ($\sim {\rm kpc}$), rather than their
density structure. As a result, 
the star formation threshold density and $\Sigma_{\rm den,gas}$ are
independent quantities. 
Consequently, $\tau_{\rm SF}$ will also be independent of
$\Sigma_{\rm gas, den}$. Therefore, variations in $\tau_{\rm SF}$ will
increase the scatter of the scaling between $\dot{\Sigma}_{\rm SF}$
and $\Sigma_{\rm gas, den}$ but it will not change its slope. As long
as the variations and the associated scatter are not so large that the
scaling is lost, the star formation law would not be affected. 

 Krumholz and Thompson (2007) have  
argued that the reason behind the Gao and Solomon (2004) findings
  is that the gas density tracer used in the particular study
 is sensitive to only a very narrow range of densities, which 
 corresponds to a specific dynamical timescale value,
while the star formation
 timescale in general does depend on the density, so sampling a wide
 density range results in the nonlinearity of the star formation
 law. 
 Although this explanation may
reconcile the traditional picture with observations of $n_{\rm gas}$
 between $1.0$ and $1.5$, it cannot account for values {\em greater}
 than $1.5$, such as those suggested by observations of dwarf
 galaxies. 

Our analysis was performed, as a first
step, for a three-dimensional cube, rather than a disc geometry which
would be more intuitive and  natural for the description of
star-forming galaxies. As a result, the length along the line of sight
of the cube fragments over which surface densities were averaged were
generally larger than their plane-of-the-sky dimensions. On the
other hand, in observations quoted here, the line-of-sight length of
the regions studied were generally comparable to their
plane-of-the-sky dimensions. We have tested that the dependence of our
scalings on the plane-of-the-sky size of the ``patches'' we have used
is minimal. However, 
we plan to return to this problem in a more detailed analysis for
different ISM geometries in the future.   

We should finally note that in our models, the multifractal 
properties are set by the mass fractions $f_i$. In turn, 
the values of $f_i$ represent a parametrization of the shape of
the multifractal PDF. The values of $f_i$ for the specific models
discussed in this work were chosen so as to represent a variety of
PDF shapes appropriate for a range of possible ISM geometries. 
The correspondence of specific PDF shapes to particular astrophysical
systems is left to be investigated through simulations and observations.

\section{Summary}\label{sum}
Observations of the scaling of $\dot{\Sigma}_{\rm SF}$ with the
surface density of gas of different local volume densities indicate
that a nonlinear scaling also exists 
between the surface densities of very dense gas
(traced e.g. by HCN) and total gas in star-forming galaxies,
$\Sigma_{\rm den, gas} \propto \Sigma_{\rm  gas}^n$. In this paper, we
have explored the possibility that this nonlinear 
scaling may arise as a result of the multifractal
topology of the ISM.
We have investigated the properties of this scaling in simulated
multifractal cubes, the geometrical properties of 
which are characterised by the values of the mass fractions $f_i$ that each
cell inherits from its parent cell and the number of levels of
refinement, $H$. Our findings can be 
summarized as follows.  

\begin{itemize}
\item A nonlinear scaling between the surface densities of very dense
  gas and total gas, $\Sigma_{\rm den, gas} \propto \Sigma_{\rm
    gas}^{n}$ is a natural property of
  multifractal geometries. The slope $n$ of the scaling generally
  depends on the minimum density of the gas that is considered as
  ``very dense'', and tends to increase as this minimum density
  increases. However, this trend does not continue indefinitely, and
  $n$ saturates to some value once the ``very dense gas'' threshold
  becomes higher than about one order of magnitude above the mean
  density. 

\item The saturation value of $n$ depends on the width of the global PDF. 
The saturation  
slope decreases as the width of the PDF increases, but again this trend is not
continued indefinitely, and the saturation slope settles to a value between 
$1.5-1.6$ for PDFs with full-width at half-maximum larger than about 2
decs in density.  The robustness of this scaling for wide multifractal PDFs
is reminiscent of the universality of the scaling slope $n_{\rm gas}$
($\approx n$, see Eq. \ref{densdens})
of the star-formation law.

\item The origin of the nonlinearity of the $\Sigma_{\rm den,gas} -
\Sigma_{\rm gas}$ scaling is the property of the
 local PDFs to extend to higher volume densities as the total surface
 density increases. This variation in spread
  between local PDFs is an intrinsic property of multifractal
  geometries. 

\item With the additional assumption that, at very high densities, 
 $\tau_{\rm SF}$ is roughly 
    constant and independent of $\Sigma_{\rm den, gas}$ 
 (consistent with observations by
  Gao and Solomon 2004 and by Wu et al.\ 2005), a geometrical
  interpretation of the   $\Sigma_{\rm den,gas} -
\Sigma_{\rm gas}$ can be extended to the star formation law
itself. Since in this case $\dot{\Sigma}_{\rm SF}\propto \Sigma_{\rm den,gas}$, we
immediately obtain $\dot{\Sigma}_{\rm SF}\propto \Sigma_{\rm
  gas}^n$. It is intriguing that in this framework two assumptions
(multifractality of the ISM and roughly constant $\tau_{\rm SF}$ at very high
densities) are sufficient to explain many different features of the
star formation law, including the robustness of the $1.5$ slope
across different systems and morphologies, the increase of the slope
in dwarf galaxies, and the decrease of the slope when tracers
sensitive to higher densities (e.g.\ CO) are used. 

\end{itemize}
\section*{Acknowledgments}
{I thank Andrey Kravtsov for discussions that inspired this work and
  for his continued encouragement and advice, and Arieh K\"{o}nigl,
  Matt Kunz, Telemachos Mouschovias, and Vasiliki Pavlidou for
  enlightening discussions and comments which have improved this paper.  
This work was supported by
NSF grants AST 02-06216 and AST02-39759, by the NASA
Theoretical Astrophysics Program grant NNG04G178G and
the Kavli Institute for Cosmological Physics through the
grant NSF PHY-0114422.}

\end{document}